\documentclass[aps,prd,twocolumn,twoside,superscriptaddress,floatfix]{revtex4-1}
\usepackage{amsmath}
\usepackage{bm}

\newcommand\nhat{\hat{\mathbf n}}

\newcommand\beq{\begin{equation}}
\newcommand\eeq{\end{equation}}
\newcommand\beqn{\begin{eqnarray}}
\newcommand\eeqn{\end{eqnarray}}

\newcommand\avesmaller[1]{\bigl\langle {#1} \bigr\rangle}

\newcommand{\ba}{\begin{eqnarray}}
\newcommand{\ea}{\end{eqnarray}}
\newcommand{\be}{\begin{equation}}
\newcommand{\ee}{\end{equation}}

\newcommand\lsim{\mathrel{\rlap{\lower4pt\hbox{\hskip1pt$\sim$}}
        \raise1pt\hbox{$<$}}}
\newcommand\gsim{\mathrel{\rlap{\lower4pt\hbox{\hskip1pt$\sim$}}
        \raise1pt\hbox{$>$}}}

\newcommand{\aaps}{{Astron.~Astrophys.~Supp.}}

\newcommand{\aap}{{Astron.~Astrophys.}}
\newcommand{\apjl}{{Astrophys.~J.~Lett.}}
\newcommand{\apjs}{{Astrophys.~J.~Supp.}}

\newcommand{\mnras}{{Mon.~Not.~R.~Astron.~Soc.}}
\usepackage{graphicx}
\usepackage[large]{subfigure}
\usepackage{amssymb, amsmath}
\usepackage[amssymb]{SIunits}
\usepackage{epstopdf}
\usepackage{aas_macros}
\usepackage{natbib}

\begin{document}

\title{The Atacama Cosmology Telescope: A Measurement of the Thermal Sunyaev-Zel'dovich Effect Using the Skewness of the CMB Temperature Distribution}
\author{Michael~J.~Wilson}
 \affiliation{Dept.~of Astrophysics, Oxford University, Oxford, 
UK OX1 3RH}
 \affiliation{Dept.~of Astrophysical Sciences, Peyton Hall, Princeton University, Princeton, NJ USA 08544}
\author{Blake D.~Sherwin}
\email{Corresponding author, email bsherwin@princeton.edu}
\affiliation{Dept.~of Physics,
Princeton University, Princeton, NJ, USA 08544}
 \author{J.~Colin Hill}
 \affiliation{Dept.~of Astrophysical Sciences, Peyton Hall, Princeton University, Princeton, NJ USA 08544}
 \author{Graeme~Addison}
 \affiliation{Dept.~of Astrophysics, Oxford University, Oxford, 
UK OX1 3RH}
 \author{Nick~Battaglia}
 \affiliation{Department of Physics, Carnegie Mellon University, Pittsburgh, PA 15213 USA}
 \author{J.~Richard~Bond}
 \affiliation{CITA, University of
Toronto, Toronto, ON, Canada M5S 3H8}
 \author{Sudeep~Das}
 \affiliation{BCCP, Dept.~of Physics, University of California, Berkeley, CA, USA 94720}\affiliation{Dept.~of Physics,
Princeton University, Princeton, NJ, USA 08544}\affiliation{Dept.~of Astrophysical Sciences, Peyton Hall, Princeton University, Princeton, NJ USA 08544}
\author{Mark~J.~Devlin}\affiliation{Dept.~of Physics and Astronomy, University of
Pennsylvania, Philadelphia, PA, USA 19104}
\author{Joanna~Dunkley}\affiliation{Dept.~of Astrophysics, Oxford University, Oxford, 
UK OX1 3RH}
\author{Rolando~D\"{u}nner}\affiliation{Departamento de Astronom{\'{i}}a y Astrof{\'{i}}sica, Pontific\'{i}a Univ. Cat\'{o}lica,
Casilla 306, Santiago 22, Chile}
\author{Joseph~W.~Fowler}\affiliation{NIST Quantum Devices Group, 325
Broadway Mailcode 817.03, Boulder, CO, USA 80305}\affiliation{Dept.~of Physics,
Princeton University, Princeton, NJ, USA 08544}
\author{Megan B.~Gralla}
\affiliation{Dept.~of Physics and Astronomy, The Johns Hopkins University, Baltimore, MD 21218-2686}
\author{Amir~Hajian}\affiliation{CITA, University of
Toronto, Toronto, ON, Canada M5S 3H8}
\author{Mark~Halpern}\affiliation{Dept.~of Physics and Astronomy, University of
British Columbia, Vancouver, BC, Canada V6T 1Z4}
\author{Matt~Hilton}
\affiliation{School of Physics \& Astronomy, University of Nottingham, Nottingham, NG7 2RD, UK}
\author{Adam~D.~Hincks}\affiliation{CITA, University of
Toronto, Toronto, ON, Canada M5S 3H8}
\author{Ren\'ee~Hlozek}\affiliation{Dept.~of Astrophysical Sciences, Peyton Hall, Princeton University, Princeton, NJ USA 08544}
\author{Kevin~Huffenberger}
\affiliation{Department of Physics, University of Miami, Coral Gables, Florida 33146}
\author{John~P.~Hughes}\affiliation{Dept.~of Physics and Astronomy, Rutgers, 
The State University of New Jersey, Piscataway, NJ USA 08854-8019}
\author{Arthur~Kosowsky}\affiliation{Dept.~of Physics and Astronomy, University of Pittsburgh, 
Pittsburgh, PA, USA 15260}
\author{Thibaut~Louis}
 \affiliation{Dept.~of Astrophysics, Oxford University, Oxford, 
UK OX1 3RH}
\author{Tobias~A.~Marriage}\affiliation{Dept.~of Physics and Astronomy, The Johns Hopkins University, Baltimore, MD 21218-2686}\affiliation{Dept.~of Astrophysical Sciences, Peyton Hall, 
Princeton University, Princeton, NJ USA 08544}
\author{Danica~Marsden}\affiliation{Department of Physics, University of California, Santa Barbara, CA 93106 US}
\author{Felipe~Menanteau}\affiliation{Dept.~of Physics and Astronomy, Rutgers, 
The State University of New Jersey, Piscataway, NJ USA 08854-8019}
\author{Kavilan~Moodley}\affiliation{Astrophysics and Cosmology Research Unit, Univ. of KwaZulu-Natal, Durban, 4041,
South Africa}
\author{Michael~D.~Niemack}\affiliation{NIST Quantum Devices Group, 325
Broadway Mailcode 817.03, Boulder, CO, USA 80305}\affiliation{Dept.~of Physics,
Princeton University, Princeton, NJ, USA 08544}
\author{Michael~R.~Nolta}\affiliation{CITA, University of
Toronto, Toronto, ON, Canada M5S 3H8}
\author{Lyman~A.~Page}\affiliation{Dept.~of Physics,
Princeton University, Princeton, NJ, USA 08544}
\author{Bruce~Partridge}\affiliation{Dept.~of Physics and Astronomy, Haverford College, Haverford, PA, USA 19041}
\author{Erik~D.~Reese}\affiliation{Dept.~of Physics and Astronomy, University of
Pennsylvania, Philadelphia, PA, USA 19104}
\author{Neelima~Sehgal}\affiliation{Dept.~of Astrophysical Sciences, Peyton Hall, Princeton University, Princeton, NJ USA 08544}
\author{Jon~Sievers}\affiliation{CITA, University of
Toronto, Toronto, ON, Canada M5S 3H8}
\author{David~N.~Spergel}\affiliation{Dept.~of Astrophysical Sciences, Peyton Hall, 
Princeton University, Princeton, NJ USA 08544}
\author{Suzanne~T.~Staggs}\affiliation{Dept.~of Physics,
Princeton University, Princeton, NJ, USA 08544}
\author{Daniel~S.~Swetz}\affiliation{Dept.~of Physics and Astronomy, University of
Pennsylvania, Philadelphia, PA, USA 19104}\affiliation{NIST Quantum Devices Group, 325
Broadway Mailcode 817.03, Boulder, CO, USA 80305}
\author{Eric~R.~Switzer}\affiliation{Kavli Institute for Cosmological Physics, 
5620 South Ellis Ave., Chicago, IL, USA 60637}\affiliation{Dept.~of Physics,
Princeton University, Princeton, NJ, USA 08544}
\author{Hy~Trac}
\affiliation{Department of Physics, Carnegie Mellon University, Pittsburgh, PA 15213 USA}
\author{Ed~Wollack}\affiliation{Code 553/665, NASA/Goddard Space Flight Center,
Greenbelt, MD, USA 20771}


\begin{abstract}
We present a detection of the unnormalized skewness $\avesmaller{\tilde
  T^3(\nhat)}$ induced by the thermal Sunyaev-Zel'dovich (tSZ) effect in
  filtered Atacama Cosmology Telescope (ACT) $148~\mathrm{GHz}$ cosmic microwave background temperature
  maps. Contamination due to infrared and radio sources is minimized by
  template subtraction of resolved sources and by constructing a mask
  using outlying values in the $218~\mathrm{GHz}$ (tSZ-null) ACT maps.
  We measure $\avesmaller{\tilde T^3(\nhat)} = -31 \pm 6 \ \mu \mathrm{K}^3$
  (Gaussian statistics assumed) or $\pm 14 \ \mu \mathrm{K}^3$ (including
  non-Gaussian corrections) in the filtered ACT data, a 5$\sigma$ detection. We show that the skewness is a sensitive probe of $\sigma_8$, and use analytic
  calculations and tSZ simulations to obtain cosmological constraints from this
  measurement. From this signal alone we infer a value of $\sigma_8=
  0.79^{+0.03}_{-0.03}$ (68\% C.L.) ${}^{+0.06}_{-0.06}$ (95\% C.L.). Our results demonstrate that
  measurements of non-Gaussianity can be a useful method for
  characterizing the tSZ effect and extracting the underlying
  cosmological information.
\end{abstract}
\maketitle

\section{Introduction}
Current observations of the cosmic microwave background (CMB)
anisotropies on arcminute scales using experiments such as the Atacama Cosmology
Telescope \citep[ACT;][]{Fow07,Swe11,Dun12} and the South
Pole Telescope \citep[SPT;][]{Car11,Sch11} probe not only the
primordial microwave background fluctuations sourced 13.7 billion
years ago, but also measure secondary anisotropies caused by more
recent and less distant physical processes. Such secondary anisotropies are
induced by infrared (IR) dusty galaxies and radio sources, gravitational
lensing, and the Sunyaev-Zel'dovich (SZ) effect. The SZ effect
\cite{Zel69,Sun70} arises due to the inverse Compton scattering of CMB
photons off high energy electrons located predominantly in hot gas in
galaxy clusters (the intra-cluster medium, or ICM). This scattering
modifies the spectrum of CMB photons in the direction of a cluster in a way that
depends on both the thermal energy contained in the ICM (the thermal
SZ effect) as well as the peculiar velocity of the cluster with
respect to the CMB rest frame (the kinetic SZ effect).  The kinetic SZ
effect simply increases or decreases the amplitude of the CMB spectrum
in the direction of a cluster, but the thermal SZ (tSZ) effect modifies the
CMB spectrum in a frequency-dependent manner.  The tSZ
effect is characterized by a decrease (increase) in the observed CMB
temperature at frequencies below (above) 218 GHz in the direction of a
galaxy cluster due to inverse Compton scattering. The thermal effect is generally at least an order of magnitude
larger than the kinetic effect for a typical massive cluster at 148 GHz.  Measurements
of the tSZ signal, which is proportional to the integrated ICM pressure along the
line of sight, can be used to observe the high redshift universe,
constrain cosmological parameters -- in particular $\sigma_8$, the variance of matter fluctuations on scales of $8~\mathrm{Mpc}/h$ -- and
probe baryonic physics in the ICM.

The tSZ signal has so far primarily been studied either by directly
resolving individual clusters in arcminute-scale CMB maps
\cite{Sta,Van,Hin,Mar,Seh} or by measuring it statistically through
its presence in the small-scale CMB power spectrum
\cite{Dun,Kei}. However, it is by no means obvious that the power
spectrum is the best way to characterize the statistical properties of
the tSZ field. Indeed, measuring the tSZ signal in the power spectrum
is challenging because there are many other sources of CMB power on
arcminute scales: primordial CMB fluctuations, CMB lensing,
instrumental noise, dusty star-forming IR galaxies, and radio
sources. In order to disentangle these contributions to the
power spectrum and isolate the amplitude of the tSZ signal, a
sophisticated multifrequency analysis is required, which involves
modeling the power spectrum contribution of each of these components
in at least two frequency bands.

In this paper we instead measure the tSZ signal using the
unnormalized skewness of the filtered temperature fluctuation
$\avesmaller{\tilde T^3(\nhat)}$. This quantity has the significant advantage
that, unlike measurements of the tSZ effect through the power
spectrum, its measurement does not require the subtraction of Gaussian
contributions, because it is only sensitive to non-Gaussian signals with non-zero skewness. The primordial CMB (which is assumed to be Gaussian on
these scales) and instrumental noise (which is Gaussian) hence do not
contribute to it. In addition, CMB lensing and the kinetic SZ effect do not
induce skewness (as they are equally likely to produce positive and negative fluctuations), and so do not
contribute either. The primary contributions to this quantity are
thus only the tSZ effect and point sources. These signals have a different frequency dependence. Furthermore, the tSZ signal contributes negative skewness, whereas radio and IR point sources
contribute positive skewness. These characteristics allow the tSZ signal to be
effectively isolated and studied, as first pointed out in \cite{Rub03}.

Measurements of the skewness also possess significant advantages from an
astrophysical perspective. A consistent problem plaguing studies of
the tSZ power spectrum has been theoretical uncertainty in the ICM
electron pressure profile \cite{Bat11,Arn10,Sha10}, especially in the
low-mass, high-redshift groups and clusters that contribute much of
the signal. As discussed in the following section in detail, the tSZ skewness signal is dominated by
characteristically higher-mass, lower-redshift clusters than those
that source the power spectrum signal.  The ICM astrophysics for
these objects is better constrained by X-ray observations and they are less sensitive to
energy input from non-gravitational sources \cite{Sha10,Bat11b}. Thus, the theoretical
systematic uncertainty in modeling the tSZ skewness is correspondingly lower as
well. In addition, at 148 GHz, dusty star-forming galaxies are less prevalent in massive,
low-redshift clusters (which contribute more to the skewness) than in high-redshift groups and clusters (which
contribute more to the tSZ power spectrum) \citep{Haj12}. Thus, we expect the
correlation between tSZ signal and dusty galaxy emission, which can complicate analyses of the tSZ effect, to be smaller for a measurement of the skewness.

Moreover, the tSZ skewness scales with a higher power of
$\sigma_8$ than the tSZ power spectrum amplitude. This result is precisely what one would expect
if the signal were dominated by higher-mass, rarer objects, as the high-mass tail of the mass function is particularly sensitive to a change in $\sigma_8$.  This provides the prospect of tight
constraints on cosmological parameters from the skewness
that are competitive with constraints from the power spectrum.

In this paper, we first explain the usefulness of the skewness as a cosmological probe by theoretically deriving its scaling with $\sigma_8$ as well as the characteristic masses of the objects sourcing the signal. Subsequent sections of the paper describe how we measured this skewness in the ACT data. We describe how the ACT temperature maps are
processed in order to make a reliable measurement of the unnormalized
skewness due to the tSZ effect, and discuss how contamination from
IR dusty galaxies and radio point sources is minimized. We
report the measurement results and discuss how the errors are
calculated. Finally, we discuss the cosmological constraints and
associated uncertainties derived from this measurement.

We assume a flat $\Lambda$CDM cosmology throughout, with parameters
set to their WMAP5 values \cite{Kom09} unless otherwise specified.  All masses
are quoted in units of $M_{\odot}/h$, where $h \equiv H_0/(100 \, \mathrm{km} \, \mathrm{s}^{-1} \, \mathrm{Mpc}^{-1})$ and $H_0$ is the Hubble parameter today.

\section{Skewness of the tSZ Effect}
In this section, we investigate 
the $N^{\mathrm{th}}$ moments of the pixel probability density function, $\avesmaller{T^N} \equiv \avesmaller{T(\nhat)^N}$, focusing on the specific case of the unnormalized skewness $\avesmaller{T^3}$.  We show that
the unnormalized skewness $\avesmaller{T^3}$ has a steeper scaling with $\sigma_8$ than the power spectrum amplitude and is dominated by characteristically higher-mass, lower-redshift clusters, for which the ICM astrophysics is better constrained and modeled. As explained earlier, these characteristics make tSZ skewness measurements a useful cosmological probe.

In order to calculate the $N^{\mathrm{th}}$ moment of the tSZ field, we assume the distribution of clusters on the sky can be adequately described by a Poisson distribution (and
that contributions due to clustering and overlapping sources are negligible \cite{Kom99}). The $N^{\mathrm{th}}$ moment is then given by
\begin{equation}
\label{eq.Npoint}
\avesmaller{T^N} = \int dz \frac{dV}{dz} \int dM \frac{dn(M,z)}{dM} \int d^2 {\bm \theta} \, T ({\bm \theta}; M,z)^N \, ,
\end{equation}
where $ T({\bm \theta}; M,z)$ is the tSZ temperature decrement at position ${\bm \theta}$
on the sky with respect to the center of a cluster of mass $M$ at redshift $z$:
\begin{eqnarray}
\label{eq.DeltaT}
T({\bm \theta}; M,z) &=& g(\nu)T_{\mathrm{CMB}}  \frac{\sigma_T}{m_e c^2} \nonumber \\&\times&\int P_e \left( \sqrt{l^2 + d_A^2(z) |{\bm \theta}|^2} ; M,z \right) dl \, ,
\end{eqnarray}
where $g(\nu)$ is the spectral function of the tSZ effect, $d_A(z)$ is the
angular diameter distance to redshift $z$, and the integral is taken over the electron pressure profile $P_e({\bm r}; M,z)$ along the
line of sight.

For a given cosmology, Eqs.~(\ref{eq.Npoint}) and (\ref{eq.DeltaT}) show that
there are two ingredients needed to calculate the $N^{\mathrm{th}}$ tSZ moment (in addition to the
comoving volume per steradian $dV/dz$, which can be calculated easily): (1) the halo mass function
$dn(M,z)/dM$ and (2) the electron pressure profile $P_e({\bm r}; M,z)$ for halos of mass $M$ at
redshift $z$.  We use the halo mass function of \cite{Tin08} with the redshift-dependent
parameters given in their Eqs.~(5)--(8).  While uncertainties in tSZ calculations due to the mass function are often neglected, they may be more important for the skewness than the power spectrum, as the skewness is more sensitive to the high-mass exponential tail of the mass function.  We estimate the uncertainty arising from the mass function by performing alternate calculations with the mass function of \cite{She02}, which predicts more massive clusters at low redshift than \cite{Tin08} for the same cosmology.  As an example, the predicted skewness calculated using the pressure profile of \cite{Bat11} with the mass function of \cite{Tin08} is $\approx 35$\% lower than the equivalent result using the mass function of \cite{She02}. However, the derived scalings of the variance and skewness with $\sigma_8$ computed using \cite{She02} are identical to those found below using \cite{Tin08}.  Thus, the scalings calculated below are robust to uncertainties in the mass function, and we use them later to interpret our skewness measurement. However, we rely on cosmological simulations to obtain predicted values of the tSZ skewness. We do not consider alternate mass functions any further in our analytic calculations.

We consider three different pressure profiles from \cite{Bat11,Arn10,Kom02} in order to evaluate the theoretical uncertainty in the scaling of the tSZ skewness with $\sigma_8$.  These profiles differ in how they are derived and in the ICM physics they assume. They thus provide a measure of the scatter in the scalings of the variance and skewness with $\sigma_8$ due to uncertainties in the gas physics.

Finally, in order to make a faithful comparison between the theory and data, we convolve Eq.~(\ref{eq.DeltaT}) with the Fourier-space filter described in the subsequent data analysis sections of this paper.  
In addition, we account for the $12\sigma$ pixel fluctuation cutoff used in the data analysis (see below) by placing each ``cluster'' of mass $M$ and redshift $z$ in the integrals of Eq.~(\ref{eq.Npoint}) in an idealized ACT pixel and computing the observed temperature decrement, accounting carefully for geometric effects that can arise depending on the alignment of the cluster and pixel centers.  If the calculated temperature decrement exceeds the $12\sigma$ cutoff, then we do not include this cluster in the integrals.  These steps cannot be neglected, as the filter and cutoff reduce the predicted tSZ skewness amplitude by up to $95$\% compared to the pure theoretical value \cite{Hil12}.  Most of this reduction is due to the filter, which modestly suppresses the temperature decrement profile of a typical cluster; this suppression strongly affects the skewness because it is a cubic statistic.


The analytic theory described above determines the scaling of the
$N^{\mathrm{th}}$ tSZ moment with $\sigma_8$.  In particular, we compute
Eq.~(\ref{eq.Npoint}) with $N=2$, $N=3$, and $N=6$ for each of the chosen
pressure profiles while varying $\sigma_8$. The scalings of the variance ($N=2$), the skewness ($N=3$), and the sixth moment ($N=6$, which we require for error calculation) with $\sigma_8$ are well-described by power laws for each of these profiles: $\avesmaller{{\tilde
    T}^{2,3,6}} \propto \sigma_8^{\alpha_{2,3,6}}$.  For the profile of
\cite{Bat11}, we find $\alpha_2 = 7.8$, $\alpha_3 = 11.1$, and $\alpha_6=16.7$; for the profile of
\cite{Arn10}, we find $\alpha_2 = 8.0$, $\alpha_3 = 11.2$, and $\alpha_6=15.9$; and for the profile of
\cite{Kom02}, we find $\alpha_2 = 7.6$, $\alpha_3 = 10.7$, and $\alpha_6=18.0$. 
Note that the scaling of the variance matches the scaling of
the tSZ power spectrum amplitude that has been found by a number of other studies, as
expected (e.g., \cite{Kom02,Tra11}).  The scaling of the unnormalized skewness is similar to that found by \cite{Hol07}, who obtained
$\alpha_3 = 10.25$.  Also, note that the skewness scaling is modified slightly from its pure theoretical value \cite{Hil12} due to the Fourier-space filter and pixel fluctuation cutoff mentioned above. The overall conclusion is that the skewness scales with a higher power of $\sigma_8$ than the variance (or power spectrum).  We use this scaling to derive a constraint on $\sigma_8$ from our measurement of the skewness below.

In addition, we compare the characteristic mass scale responsible for the tSZ
skewness and tSZ variance (or power spectrum) signals.  Analytic
calculations show that the tSZ power spectrum amplitude typically
receives $\approx 50$\% of its value from halos with $M < 2$--$3
\times 10^{14} \,\, \mathrm{M}_{\odot}/h$, while the tSZ skewness
receives only $\approx 20$\% of its amplitude from these less massive
objects.  This indicates that the clusters responsible for the tSZ skewness signal are better theoretically modeled than those responsible for much of the tSZ power spectrum, both because massive clusters have been observed more thoroughly, and because more massive clusters are dominated
by gravitational heating and are less sensitive to non-linear energy input from active galactic nuclei, turbulence,
and other mechanisms \cite{Sha10,Bat11b}. We verify this claim when interpreting the skewness measurement below, finding that the systematic theoretical uncertainty (as derived from simulations) is slightly smaller than the statistical error from the measurement, though still non-negligible. 

\section{Map Processing}
\label{processing1}
\subsection{Filtering the Maps}
The Atacama Cosmology Telescope \cite{Fow07,Swe11,Dun12} is a 6m telescope in the Atacama Desert of Chile, which operated at 148, 218, and 277 GHz using three 1024-element arrays of superconducting bolometers. The maps used in this analysis were made over three years of observation in the equatorial region during 2008--2010 at 148 GHz, and consist of six $3\degree \times 18\degree$ patches of sky at a noise level of $\approx 21\ \mu \mathrm{K}$ arcmin. In our source mask construction we also use maps of the same area made in 2008 at a frequency of 218 GHz. The maps were calibrated as in \cite{HajCal}. We apodize the maps by multiplying them with a mask that smoothly increases from zero to unity over $0.1\degree$ from the edge of the maps.

Although atmospheric noise is removed in the map-making process, we
implement an additional filter in Fourier space to remove signal at multipoles below $\ell=500$ ($\ell$ is the magnitude of the Fourier variable conjugate to sky angle). In addition, we remove a stripe for which $-100< \ell_{\mathrm{dec}}<100$ along the Fourier axis corresponding to declination to avoid contamination by scan noise. Furthermore, to increase the tSZ
signal-to-noise, we apply a Wiener filter which downweights scales at
which the tSZ signal is subdominant. This (non-optimal) filter is
constructed by dividing the best-fit tSZ power spectrum
from \cite{Dun} by the total average power spectrum measured in the
data maps, i.e. $C_{\ell}^{tSZ}/C_\ell^{tot}$. For multipoles above
$\ell = 6 \times 10^{3}$, the tSZ signal is completely dominated by
detector noise and point sources, and hence we remove all power above
this multipole in the temperature maps. The final Fourier-space
filter, shown in Fig.~1, is normalized such that its
maximum value is unity. As it is constructed using the binned power spectrum of the data, it is not perfectly smooth; however, we apply the same filter consistently to data, simulations, and analytic theory, and thus any details of the filter do not bias the interpretation of our result. After filtering, the edges of the maps are cut off to reduce any edge effects that might occur upon Fourier transforming despite apodization. Simulations verify that no additional skewness is introduced by edge effects into a trimmed map.

\subsection{Removing Point Sources}
\label{processing2}
In order to obtain a skewness signal due only to the tSZ effect,
any contamination of the signal by point sources must be minimized. These objects consist of IR dusty galaxies and radio sources. We use two approaches to eliminate the point source contribution: template subtraction and masking using the 218 GHz channel.

\begin{figure}[!h]
\label{fig.filter}
  \begin{center}
    \includegraphics[width=\columnwidth]{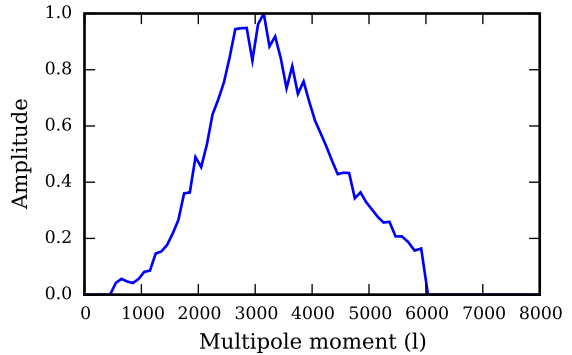}
    \caption{The Wiener filter applied to the ACT temperature maps
    before calculating the unnormalized skewness. This filter
    upweights scales on which the tSZ signal is large compared to
    other sources of anisotropy.}
  \end{center}
\end{figure}

In the template subtraction method \cite{Mar11}, which we use to remove resolved point sources (mainly bright radio sources), sources with a signal-to-noise (S/N) greater than five are first identified in a match-filtered map. A template with the shape of the ACT beam is then scaled to the appropriate peak brightness of each source, and this profile is subtracted from the raw data. The process is iterated following the CLEAN algorithm \cite{Hog74} until no more sources can be identified. We verify that this procedure does not introduce skewness into the maps (e.g., through oversubtraction) by checking that similar results are obtained using a different procedure for reducing source contamination, in which we mask and in-paint pixels which contain bright sources with S/N $>5$ \cite{thibaut}.

We take a second step to suppress the lower-flux, unresolved point sources (mainly dusty galaxies) that remain undetected by the template subtraction algorithm.  At 218 GHz, dusty galaxies are significantly brighter than at 148 GHz and the tSZ effect is negligible. We construct a dusty galaxy mask by setting all pixels (which are approximately $0.25~\mathrm{arcmin}^2$) to zero that have a temperature in the 218 GHz maps larger than a specified cutoff value. This cutoff is chosen to be 3.2 times the standard deviation of the pixel values in the filtered 148 GHz map ($3.2\sigma$). This procedure ensures regions with high flux from dusty galaxies are masked. We also set to zero all pixels for which the temperature is lower than the negative of this cutoff, so that the masking procedure does not introduce spurious skewness into the lensed CMB distribution, which is assumed to have zero intrinsic skewness. The mask is then applied to the 148 GHz map to reduce the point source contribution. Simulations (\cite{Seh10} for IR sources) verify that the masking procedure does not introduce spurious skewness into the 148 GHz maps.

\begin{figure}[!h]
\label{fig.pdf}
  \begin{center}
    \includegraphics[width=\columnwidth]{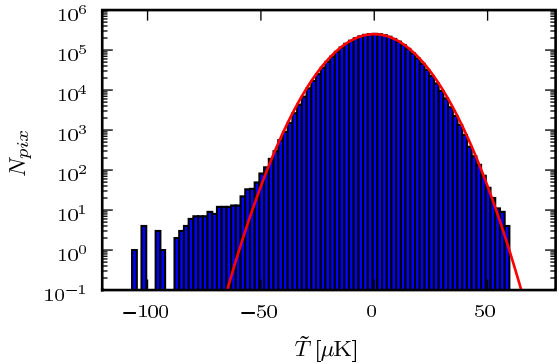}
    \caption{Histogram of the pixel temperature values in the filtered, masked ACT CMB temperature maps. A Gaussian curve is overlaid in red.}
  \end{center}
\end{figure}

Finally, all pixels more than twelve standard deviations ($12\sigma$) from the mean are also removed from the 148 GHz maps.  Due to the ringing around very positive or negative pixels caused by the Wiener filter, the surrounding eight arcminutes of these points are also masked.  This additional step slightly increases the S/N of the skewness measurement by reducing the dependence on large outliers, enhances the information content of low moments by truncating the tail of the pixel probability density function, and ensures that any anomalous outlying points from possibly mis-subtracted bright radio sources do not contribute to the skewness signal.  Overall, 14.5\% of the 148 GHz map is removed by the masking procedure, though the removed points are random with respect to the tSZ field and should not change the signal.

\section{Results}
\subsection{Evaluating the Skewness}
We compute the unnormalized skewness of the filtered and processed 148 GHz maps by simply cubing and averaging the pixel values in real space. The result is $\avesmaller{\tilde T^3} = -31 \pm 6 \ \mu \mathrm{K}^3$, a $5\sigma$ deviation from the null result expected for a signal without any non-Gaussian components. The skewness of the CMB temperature distribution in our filtered, processed maps is visible in the pixel value histogram shown in Fig.~2 (along with a Gaussian curve overlaid for comparison). 
It is evident that the Gaussian CMB has been recovered on the positive side by point source masking, with the apparent truncation beyond $50 \ \mu \mathrm{K}$ due to the minute probability of such temperatures in the Gaussian distribution.  The likelihood corresponding to our measurement of the skewness is shown in Fig.~3.

\begin{figure}[!h]
\label{fig.likelihood}
  \begin{center}
    \includegraphics[width=\columnwidth]{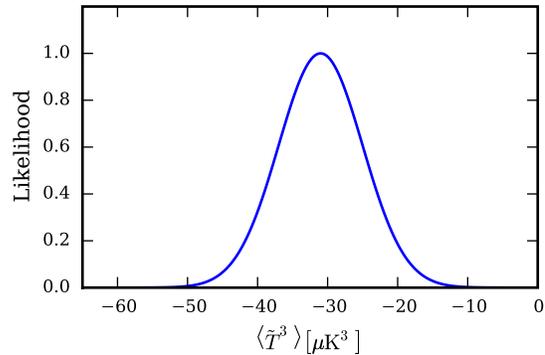}
    \caption{Likelihood of the skewness measurement described in the text 
(with Gaussian statistics assumed).}
  \end{center}
\end{figure}

The 
``Gaussian statistics assumed'' error on the skewness includes only Gaussian sources of noise.  We calculate this error by using map simulations that consist of Gaussian random fields with the same power spectrum as that observed in the data, including beam effects. 
These simulations contain Gaussian contributions from IR, SZ, and radio sources, the primordial lensed CMB, and detector noise.
This estimate thus does not include the error resulting from 
non-Gaussian corrections, which (after source subtraction) are due to the non-Gaussian tSZ signal. Though CMB lensing is also a non-Gaussian effect it does not contribute to the error on the skewness, as the connected part of the six-point function is zero to lowest order in the lensing potential, and the connected part of the three-point function is also negligibly small (see \cite{lew06} and references therein).

We calculate errors that include 
non-Gaussian corrections by constructing more realistic simulations. To construct such simulations, we add maps with simulated tSZ signal from \cite{Bat10}, which assume $\sigma_8=0.8$, to realizations of a Gaussian random field which has a spectrum such that the power spectrum of the combined map matches that observed in the ACT temperature data. Given the simulated sky area, we obtain 39 
statistically independent simulated maps, each of size $148~\mathrm{deg}^2$. 
 By applying an identical procedure to the simulations as to the data, measuring the scatter amongst the patches, and appropriately scaling the error to match the $237~\mathrm{deg}^2$ of unmasked sky in the processed map, we obtain a full error 
(including non-Gaussian corrections) on the unnormalized skewness  of $14 \ \mu \mathrm{K}^3$.
While this error is a robust estimate it should be noted that the ``error on the error'' is not insignificant due to the moderate simulated volume available.  The scatter of skewness values measured from each of the simulated maps is consistent with a Gaussian distribution.  The estimate for the full error is coincidentally the same as the standard error, $14 \ \mu \mathrm{K}^3$, estimated from the six patches into which the data are divided.  The full error is used below in deriving cosmological constraints from the skewness measurement.

\begin{figure}[!h]
  \begin{center}
    \includegraphics[width=\columnwidth]{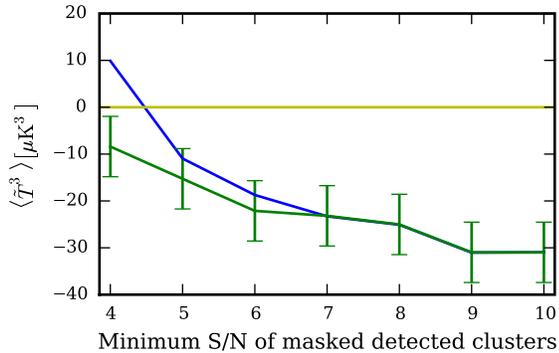}
    \caption{Plot of the skewness signal as a function of the minimum S/N of the clusters that are masked (this indicates how many known clusters are left in the data, unmasked). The blue line is calculated using the full cluster candidate catalog obtained via matched filtering, while the green line uses a catalog containing only optically-confirmed clusters \cite{Men12}.  Both lines have identical errors, but we only plot them for the green line for clarity. Confirmed clusters source approximately two-thirds of the signal, which provides strong evidence that it is due to the tSZ effect. Note that one expects a positive bias of $\approx 4 \ \mu \mathrm{K}^3$ for the S/N $=4$ point of the blue line due to impurities in the full candidate catalog masking the tail of the Gaussian distribution.}
  \end{center}
\end{figure}

\subsection{The Origin of the Signal}
Is the skewness dominated by massive clusters with large tSZ decrements -- as suggested by theoretical considerations described earlier -- or by more numerous, less massive clusters? To investigate this question, we mask clusters in our data which were found in the 148 GHz maps using a matched filter as in \cite{Mar09}. All clusters detected above a threshold significance value are masked; we vary this threshold and measure the remaining skewness in order to determine the origin of the signal.

Fig.~4 shows a plot of the signal against the cluster detection significance cutoff.  We include calculations using both the full cluster candidate catalog obtained via matched filtering as well a catalog containing only clusters confirmed optically using the methodology of \cite{Men10} on the SDSS Stripe 82 \cite{Men12}.  The SDSS Stripe 82 imaging data cover $\approx 80$\% of the total map area, and thus some skewness signal will necessarily arise from objects not accounted for in this catalog.  The results for these two catalogs agree when masking clusters with S/N $\geq 7$, but differ slightly when masking lower S/N clusters.  This effect is likely due to the small shortfall in optical follow-up area as well as a small number of false detections (i.e., impurity) in the candidate clusters that have not yet been optically followed up.

\begin{figure}[!h]
  \begin{center}
    \includegraphics[width=\columnwidth]{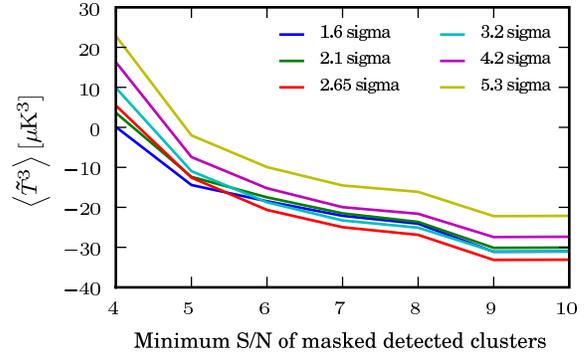}
    \caption{A test for IR source contamination: similar to the blue line in Fig.~4, but with a range of values of the cutoff used to construct an IR source mask in the 218 GHz band. Any cutoff below $\approx3.2\sigma$ gives similarly negative results and thus appears sufficient for point source removal, where $\sigma = 10.3 \ \mu \mathrm{K}$ is the standard deviation of the 148 GHz maps.  For comparison, the standard deviation of the 218 GHz maps is $\approx 2.2$ times larger.  The percentages of the map which are removed for the masking levels shown, from the least to the most strict cut, are 0.7, 2.5, 8.4, 14.5, 23.7, and 36.6\%.}
  \end{center}
\end{figure}

Using either catalog, Fig.~4 implies that just under half of the tSZ skewness is obtained from clusters that lie below a $5\sigma$ cluster detection significance, while the remainder comes from the brightest and most massive clusters. The results of \cite{Mar09} suggest that clusters detected at $5\sigma$ significance are roughly characterized by a mass $M_{500} = 5 \times 10^{14} M_{\odot}/h$, where $M_{500}$ is the mass enclosed within a radius such that the mean enclosed density is 500 times the critical density at the cluster redshift. This value corresponds to a virial mass of roughly $M = 9 \times 10^{14} M_{\odot}/h$, which was also found to be the mass detection threshold for high-significance ACT clusters in \cite{Men10}.  Fig.~4 thus demonstrates that roughly half of the tSZ skewness signal is due to massive clusters with $M \gtrsim10^{15} M_{\odot}/h$.  The theoretical calculations described earlier give similar results for the fraction of the signal coming from clusters above and below this mass scale, which is significantly higher than the characteristic mass scale responsible for the tSZ power spectrum signal.

Finally, the positive value in the full candidate catalog line shown in Fig.~4 when masking clusters above S/N $= 4$ is consistent with zero. When masking at this level (with the candidate catalog which contains some impurities), we slightly cut into the negative pixel values in the Gaussian component of Fig.~2, leading to a small spurious positive skewness.  For the points we plot, we calculate that this bias is only non-negligible for the S/N $= 4$ cut, where it is $\approx 4 \ \mu \mathrm{K}^3$. This bias effectively explains the small positive offset seen in Fig.~4.  However, we discuss positive skewness due to any possible residual point source contamination below. Overall, the dependence of the measured skewness on cluster masking shown in Fig.~4 provides strong evidence that it is caused by the tSZ effect.

\subsection{Testing for Systematic Infrared Source Contamination}
Despite our efforts to remove point sources, a small residual point source contamination of the signal could remain, leading to an underestimate of the amplitude of the tSZ skewness. To investigate this systematic error source, we vary the level at which point sources are masked in the 218 GHz maps (the original level is 3.2 times the standard deviation of the pixel values in the filtered 148 GHz map ($3.2\sigma$), as described above). The results of this test are shown in Fig.~5, which uses the full catalog of cluster candidates as described in Fig.~4, since the optically-confirmed catalog does not yet cover the entire ACT map.  
Note that masking at $3.2\sigma$ results in a skewness measurement which agrees with its apparent asymptotic limit as the IR contamination is reduced, within the expected fluctuations due to masking.  While a slightly more negative skewness value can be measured for some masking levels stricter than the $3.2 \sigma$ level chosen in the analysis, fluctuations upon changing the unmasked area of the sky are expected, so that it can not be rigorously inferred that IR contamination is reduced between $3.2\sigma$ and $2.65\sigma$. Fig.~5 suggests that masking at the $3.2\sigma$ level sufficiently removes any contamination by IR sources, and stricter cuts will reduce the map area and increase statistical errors unnecessarily. 

However, to further estimate the residual point source contamination in the 148 GHz maps, we process simulations of IR sources from \cite{Seh10} (with source amplitudes scaled down by 1.7 to match recent observations, as in 
\cite{vanE}) with the same masking procedure as that applied to the data (described in $\S$\ref{processing2}), creating a mask in a simulated 218 GHz map, and applying it to a simulated IR source signal at 148 GHz.  We find a residual signal of 
$\avesmaller{\tilde T^3} = 3.9 \pm 0.1 \ \mu \mathrm{K}^3$.  We treat this result as a bias in deriving cosmological constraints from the tSZ skewness in the following section.


We also investigate a linear combination of the 148 and 218 GHz maps that should have minimal IR source levels, namely, an appropriately scaled 218 GHz map subtracted from a 148 GHz map.  Assuming that the spatial distribution of the point sources is not affected by the difference in observation frequency between 148 and 218 GHz and a single spectral index can be applied to all sources, a simple factor of $\approx 3.2$ \cite{Dun} relates a point source's signal in the two different frequency bands.  We find that the appropriate linear combination (subtracting 1/3.2 times the 220 GHz map from the 148 GHz map) produces a signal in agreement with that resulting from the previously described masking procedure, although the additional noise present in the 218 GHz maps slightly reduces the significance of the detection.  

\section{Cosmological Interpretation}
\label{interpretation}
To obtain cosmological information from the measured amplitude of the unnormalized skewness, we compare our results with two different sets of tSZ simulations \cite{Seh10,Bat10}. Both sets of simulations are run with $\sigma_8=0.8$, but differ in their treatment of the ICM.  The simulation of \cite{Bat10} is a fully hydrodynamic cosmological simulation that includes sub-grid prescriptions for feedback from active galactic nuclei, star formation, and radiative cooling.  The simulation also captures non-thermal pressure support due to turbulence and other effects, which significantly alters the ICM pressure profile.  The simulation of \cite{Seh10} is a large dark matter-only $N$-body simulation that is post-processed to include gas according to a polytropic prescription.  This simulation also accounts for non-thermal pressure support (though with a smaller amount than \cite{Bat10}), and matches the low-redshift X-ray data presented in \cite{Arn10}.

We perform the same filtering and masking as that applied to the data in order to analyze the simulation maps. For both simulations, the filtering reduces the signal by $\approx 95$\% compared to the unfiltered value. For the simulations of \cite{Bat10}, we measure $\avesmaller{\tilde T^3}^S = -37 \ \mu \mathrm{K}^3$, with negligible errors (the superscript $S$ indicates a simulated value). However, this value is complicated by the fact that these simulations only include halos below $z=1$. An analytic estimate for the skewness contribution due to halos with $z>1$ from Eq.~(\ref{eq.Npoint}) gives a $6$\% correction, which yields $\avesmaller{\tilde T^3}^S = -39 \ \mu \mathrm{K}^3$. For the simulations of \cite{Seh10}, we measure $\avesmaller{\tilde T^3}^S = -50  \ \mu \mathrm{K}^3$, with errors also negligible for the purposes of cosmological constraints.

We combine these simulation results with our calculated scalings of the skewness and the sixth moment with $\sigma_8$ to construct a likelihood:
\beq
\label{eq.likelihood}
\mathcal{L}(\sigma_8) = \exp \left(-\frac{\left(\avesmaller{\tilde T^3}^D-\avesmaller{\tilde T^3}^{\mathrm{th}}(\sigma_8)\right)^2}{2\sigma_{\mathrm{th}}^2(\sigma_8)}\right)
\eeq
where $\avesmaller{\tilde T^3}^D$ is our measured skewness value and the theoretically expected skewness as a function of $\sigma_8$ is given by
\beq
\avesmaller{\tilde{T}^3}^{\mathrm{th}}(\sigma_8) = \avesmaller{\tilde{T}^3}^{S}  \left(\frac{\sigma_8}{0.8}\right)^{\alpha_3} \,.
\eeq
The likelihood in Eq.~(\ref{eq.likelihood}) explicitly accounts for the fact that $\sigma_{\mathrm{th}}^2$, the variance of the skewness, depends on $\sigma_8$ --- a larger value of $\sigma_8$ leads to a larger expected variance in the tSZ skewness signal.  In particular, the variance of the tSZ skewness is described by a sixth moment, so it scales as $\sigma_8^{\alpha_6}$.  As determined above, the Gaussian and non-Gaussian errors on the skewness are $6 \ \mu \mathrm{K}^3$ and $\sqrt{14^2-6^2} \ \mu \mathrm{K}^3 = 12.6 \ \mu \mathrm{K}^3$, respectively.  We approximate the dependence of the full error on $\sigma_8$ by assuming that only the non-Gaussian component scales with $\sigma_8$; while this is not exact, as some of the Gaussian error should also scale with $\sigma_8$, small differences in the size or scaling of the Gaussian error component cause negligible changes in our constraints on $\sigma_8$.

Finally, although we have argued previously that IR source contamination is essentially negligible, we explicitly correct for the residual bias as calculated in the previous section.  Thus, we replace $\avesmaller{\tilde T^3}^D = -31 \ \mu \mathrm{K}^3$ with $\avesmaller{\tilde T^3}^D_{corr} = -31 - 3.9 \ \mu \mathrm{K}^3$ in Eq.~(\ref{eq.likelihood}).  (Note that this bias correction only shifts the central value derived for $\sigma_8$ below by roughly one-fifth of the $1\sigma$ confidence interval.)  Moreover, in order to be as conservative as possible, we also model the effect of residual point sources by including an additional IR contamination error (with the same value as the residual IR source contamination, $3.9 \ \mu \mathrm{K}^3$) in our expression for the variance of the skewness:
\beq
\sigma_{\mathrm{th}}^2(\sigma_8) = 6^2 \ \mu \mathrm{K}^6 + 12.6^2 \left(\frac{\sigma_8}{0.8}\right)^{\alpha_6} \ \mu \mathrm{K}^6 + 3.9^2 \ \mu \mathrm{K}^6.
\eeq

Using the likelihood in Eq.~(\ref{eq.likelihood}), we obtain confidence intervals and derive a constraint on $\sigma_8$. Our likelihood and hence our constraints depend in principle on which simulation we use to calculate $\avesmaller{\tilde{T}^3}^{S}$, as well as on the values we choose for $\alpha_3$ and $\alpha_6$. Using the simulations of \cite{Bat10} and the scalings determined above for the profile from \cite{Bat11}, we find $\sigma_8 = 0.79^{+0.03}_{-0.03}$ (68\% C.L.) ${}^{+0.06}_{-0.06}$ (95\% C.L.).  In Table I, we compare the constraints on $\sigma_8$ obtained from the use of different scalings and simulated skewness values; the constraints are insensitive to both the pressure profile used to derive the scaling laws and the choice of simulation used to compute the skewness.

For comparison, the final release from the \emph{Chandra} Cluster Cosmology Project found $\sigma_8 = 0.803 \pm 0.0105$, assuming $\Omega_m = 0.25$ (there is a strong degeneracy between $\sigma_8$ and $\Omega_m$ for X-ray cluster measurements that probe the mass function) \cite{Vik}. Perhaps more directly comparable, recent studies of the tSZ power spectrum have found $\sigma_8 = 0.77 \pm 0.04$ (statistical error only) \cite{Dun} and $\sigma_8 = 0.807 \pm 0.016$ (statistical error and approximately estimated systematic error due to theoretical uncertainty) \cite{Rei}.  Our results are also comparable to recent constraints using number counts of SZ-detected clusters from ACT and SPT, which found $\sigma_8 = 0.851 \pm 0.115$ (fully marginalizing over uncertainties in the mass-SZ flux scaling relation) \cite{Seh11} and $\sigma_8 = 0.807 \pm 0.027$ (marginalizing over uncertainties in an X-ray-based mass-SZ flux scaling relation) \cite{Rei12}, respectively.  Although more than half of the tSZ skewness signal that we measure is sourced by detected clusters (i.e., the same objects used in the number counts analyses), our method also utilizes cosmological information from clusters that lie below the individual detection threshold, which gives it additional statistical power.  Finally, note that we have fixed all other cosmological parameters in this analysis, as $\sigma_8$ is by far the dominant parameter for the tSZ skewness \cite{Batt}.  However, marginalizing over other parameters will slightly increase our errors.

To evaluate the theoretical systematic uncertainty in the amplitude of the filtered skewness due to unknown ICM astrophysics, we test the effect of different gas prescriptions by analyzing simulations from \cite{Bat10} with all forms of feedback, radiative cooling and star formation switched off, leading to an adiabatic ICM gas model. For these adiabatic simulations we find $\avesmaller{\tilde T^3}^S = -56 \ \mu \mathrm{K}^3$ (after applying the $6\%$ correction mentioned earlier), which for the skewness we measure in our data would imply $\sigma_8 = 0.77^{+0.02}_{-0.02}$ (68\% C.L.)${}^{+0.05}_{-0.05}$ (95\% C.L.). Turning off feedback and all sub-grid physics is a rather extreme case, so the systematic theoretical uncertainty
\begin{table}[h]
\begin{tabular}{l | c | c}
\ & Battaglia  $\avesmaller{\tilde{T}^3}^{S}$  & Sehgal $\avesmaller{\tilde{T}^3}^{S}$ \\
\hline
Battaglia $\alpha_3, \alpha_6$ & $0.79^{+0.03}_{-0.03}$ ${}^{+0.06}_{-0.06}$ & $0.77^{+0.03}_{-0.02}$ ${}^{+0.05}_{-0.05}$ \\
Arnaud $\alpha_3, \alpha_6$ &$0.79^{+0.03}_{-0.03}$ ${}^{+0.06}_{-0.06}$& $0.77^{+0.02}_{-0.02}$ ${}^{+0.05}_{-0.05}$ \\
K-S $\alpha_3, \alpha_6$ & $0.79^{+0.03}_{-0.03}$ ${}^{+0.07}_{-0.06}$ & $0.77^{+0.03}_{-0.03}$ ${}^{+0.06}_{-0.05}$ \\
\end{tabular}

\caption{Constraints on $\sigma_8$ derived from our skewness measurement using two different simulations and three different scalings of the skewness and its variance with $\sigma_8$.  The top row lists the simulations used to calculate the expected skewness for $\sigma_8 = 0.8$~\cite{Bat10,Seh10}; the left column lists the pressure profiles used to calculate the scaling of the skewness and its variance with $\sigma_8$~\cite{Bat11,Arn10, Kom02}.  The errors on $\sigma_8$ shown are the 68\% and 95\% confidence levels.}
\end{table}
for a typical simulation with some form of feedback should be slightly smaller than the statistical error from the measurement, though still non-negligible. This contrasts with measurements of $\sigma_8$ via the tSZ power spectrum, for which the theoretical systematic uncertainty is comparable to or greater than the statistical uncertainty \cite{Dun,Rei}. As highlighted earlier, this difference can be traced to the dependence of the power spectrum amplitude on the ICM astrophysics within low-mass, high-redshift clusters.  The skewness, on the other hand, is dominated by more massive, lower-redshift clusters that are less affected by uncertain non-gravitational feedback mechanisms and are more precisely constrained by observations. Nonetheless, as the statistical uncertainty decreases on future measurements of the tSZ skewness, the theoretical systematic error will quickly become comparable, and thus additional study of the ICM electron pressure profile will be very useful.

\section{Conclusions}
As the thermal Sunyaev-Zel'dovich field is highly non-Gaussian, measurements of non-Gaussian signatures such as the skewness can provide cosmological constraints that are competitive with power spectrum measurements.  We have presented a first measurement of the unnormalized skewness $\avesmaller{\tilde T^3(\nhat)}$ in ACT CMB maps filtered for high signal to noise. As this is a purely non-Gaussian signature, primordial CMB and instrumental noise cannot be confused with or bias the signal, unlike measurements of the tSZ power spectrum. We measure the skewness at $5\sigma$ significance: $\avesmaller{\tilde T^3(\nhat)} = -31 \pm 6 \ \mu \mathrm{K}^3$ 
(Gaussian statistics assumed).  Including non-Gaussian corrections increases the error to $\pm \ 14 \ \mu \mathrm{K}^3$. Using analytic calculations and simulations to translate this measurement into constraints on cosmological parameters, we find $\sigma_8=
  0.79^{+0.03}_{-0.03}$ (68\% C.L.) ${}^{+0.06}_{-0.06}$ (95\% C.L.), with a slightly smaller but non-negligible systematic error due to theoretical uncertainty in the ICM astrophysics. This detection represents the first realization of a new, independent method to measure $\sigma_8$ based on the tSZ skewness, which has different systematic errors than several other common methods. With larger maps and lower noise, tSZ skewness measurements promise significantly tighter cosmological constraints in the near future.

\begin{acknowledgments}
As this manuscript was being prepared, we learned of related theoretical work by the authors of \cite{Batt}, and we acknowledge very helpful discussions with the members of this collaboration. 
This work was supported by the U.S.\ NSF through awards AST-0408698, PHY-0355328, AST-0707731 and PIRE-0507768, as well as by Princeton Univ., the Univ. of Pennsylvania, FONDAP, Basal, Centre AIUC, RCUK Fellowship (JD),  NASA grant NNX08AH30G (SD, AH, TM), NSERC PGSD (ADH), NSF AST-0546035 and AST-0807790 (AK), NSF PFC grant PHY-0114422 (ES),  KICP Fellowship (ES), SLAC no. DE-AC3-76SF00515 (NS), ERC grant 259505 (JD), BCCP (SD), and the NSF GRFP (BDS, BLS). We thank  B.\ Berger, R.\ Escribano, T.\ Evans, D.\ Faber, P.\ Gallardo, A.\ Gomez, M.\ Gordon, D.\ Holtz, M.\ McLaren, W.\ Page, R.\ Plimpton, D.\ Sanchez, O.\ Stryzak, M.\ Uehara, and Astro-Norte for assistance with ACT. ACT operates in the Parque Astron\'{o}mico Atacama in northern Chile under the auspices of Programa de Astronom\'{i}a, a program of the Comisi\'{o}n Nacional de Investigaci\'{o}n Cient\'{i}fica y Tecnol\'{o}gica de Chile (CONICYT).
\end{acknowledgments}

\end{document}